\documentclass[aps,prl, amsmath, superscriptaddress, twocolumn,sort&compress,floatfix, amssymb]{revtex4}
\pdfoutput=1
\usepackage{graphicx}
\usepackage{rotating}
\usepackage{dcolumn}
\usepackage{bm}
\usepackage[dvipsnames]{xcolor}
\usepackage[latin1]{inputenc}
\usepackage{times}
\usepackage{braket}
\usepackage[colorlinks=true,linkcolor=blue,urlcolor=blue,citecolor=blue]{hyperref}

\usepackage{multirow}

\newcommand{\nag}{{\phantom{\dagger}}}

\newcommand{\eq}[1]{Eq.\thinspace{}(\ref{#1})}
\newcommand{\eqw}[1]{(\ref{#1})}
\newcommand{\fig}[1]{Fig.\thinspace{}\ref{#1}}
\newcommand{\fc}[1]{{#1}}
\newcommand{\figc}[2]{Fig.\thinspace{}\ref{#1}\fc{#2}}

\begin{document}

\author{F. Meinert}
\affiliation{Institut f\"ur Experimentalphysik und Zentrum f\"ur Quantenphysik, Universit\"at Innsbruck, 6020 Innsbruck, Austria}
\author{M. Knap}
\affiliation{Department of Physics, Walter Schottky Institute, and Institute for Advanced Study, Technical University of Munich, 85748 Garching, Germany}
\author{E. Kirilov}
\affiliation{Institut f\"ur Experimentalphysik und Zentrum f\"ur Quantenphysik, Universit\"at Innsbruck, 6020 Innsbruck, Austria}
\author{K. Jag-Lauber}
\affiliation{Institut f\"ur Experimentalphysik und Zentrum f\"ur Quantenphysik, Universit\"at Innsbruck, 6020 Innsbruck, Austria}
\author{M. B. Zvonarev}
\affiliation{LPTMS, CNRS, Univ. Paris-Sud, Universit\'e Paris-Saclay, 91405 Orsay, France}
\author{E. Demler}
\affiliation{Department of Physics, Harvard University, Cambridge, Massachusetts 02138, USA}
\author{H.-C. N\"agerl}
\affiliation{Institut f\"ur Experimentalphysik und Zentrum f\"ur Quantenphysik, Universit\"at Innsbruck, 6020 Innsbruck, Austria}

\title{Bloch oscillations in the absence of a lattice}

\date{\today}

\begin{abstract}

We experimentally investigate the quantum motion of an impurity atom that is immersed in a strongly interacting one-dimensional Bose liquid and is subject to an external force. We find that the momentum distribution of the impurity exhibits characteristic Bragg reflections at the edge of an emergent Brillouin zone. While Bragg reflections are typically associated with lattice structures, in our strongly correlated quantum liquid they result from the interplay of short-range crystalline order and kinematic constraints on the many-body scattering processes in the one-dimensional system. As a consequence, the impurity exhibits periodic dynamics that we interpret as Bloch oscillations, which arise even though the quantum liquid is translationally invariant. Our observations are supported by large-scale numerical simulations.

\end{abstract}

\maketitle

A skydiver accelerated by the gravitational force approaches a constant drift velocity due to friction with the surrounding medium. In the quantum realm, dynamics can be significantly richer. For example, a quantum particle accelerated in a periodic crystal potential does not move at all on average but rather undergoes a periodic motion known as Bloch oscillations~\cite{BLOCH28,ZENER34}. Such an oscillatory motion is a direct consequence of the periodic momentum-dependence of the eigenstates in a lattice potential and arises from continuous translational symmetry breaking. Bloch oscillations have been observed for electrons in solid state systems~\cite{FELDMANN92} and have been investigated in detail with ultracold atoms in optical lattices~\cite{DAHAN96,Anderson98,GUSTAVSSON08,Fattori08,Meinert14}. One might expect that a quantum liquid, which is fully translational invariant, or in other words does not have an imprinted lattice structure, would preclude such striking dynamics. However, recent theoretical studies \cite{Gangardt09,Schecter12} suggest that Bloch oscillations can emerge also in the presence of a continuous translational symmetry. In particular, for impurity atoms immersed in one-dimensional (1D) quantum liquids such dynamics is expected to arise due to strong quantum correlations, which lead to effective crystal-like properties. Yet, suitable conditions for that phenomenon are debated~\cite{Gamayun14}. Ultracold quantum gases provide an ideal setting to experimentally study the dynamics of impurity particles coupled to host environments \cite{Fukuhara13,Fukuhara13b,Palzer09,Catani12,Cetina16} due to excellent parameter control, precise initial state preparation, and decoupling from the environment.

\onecolumngrid

\begin{figure}[h]
\includegraphics[width=0.373\textwidth]{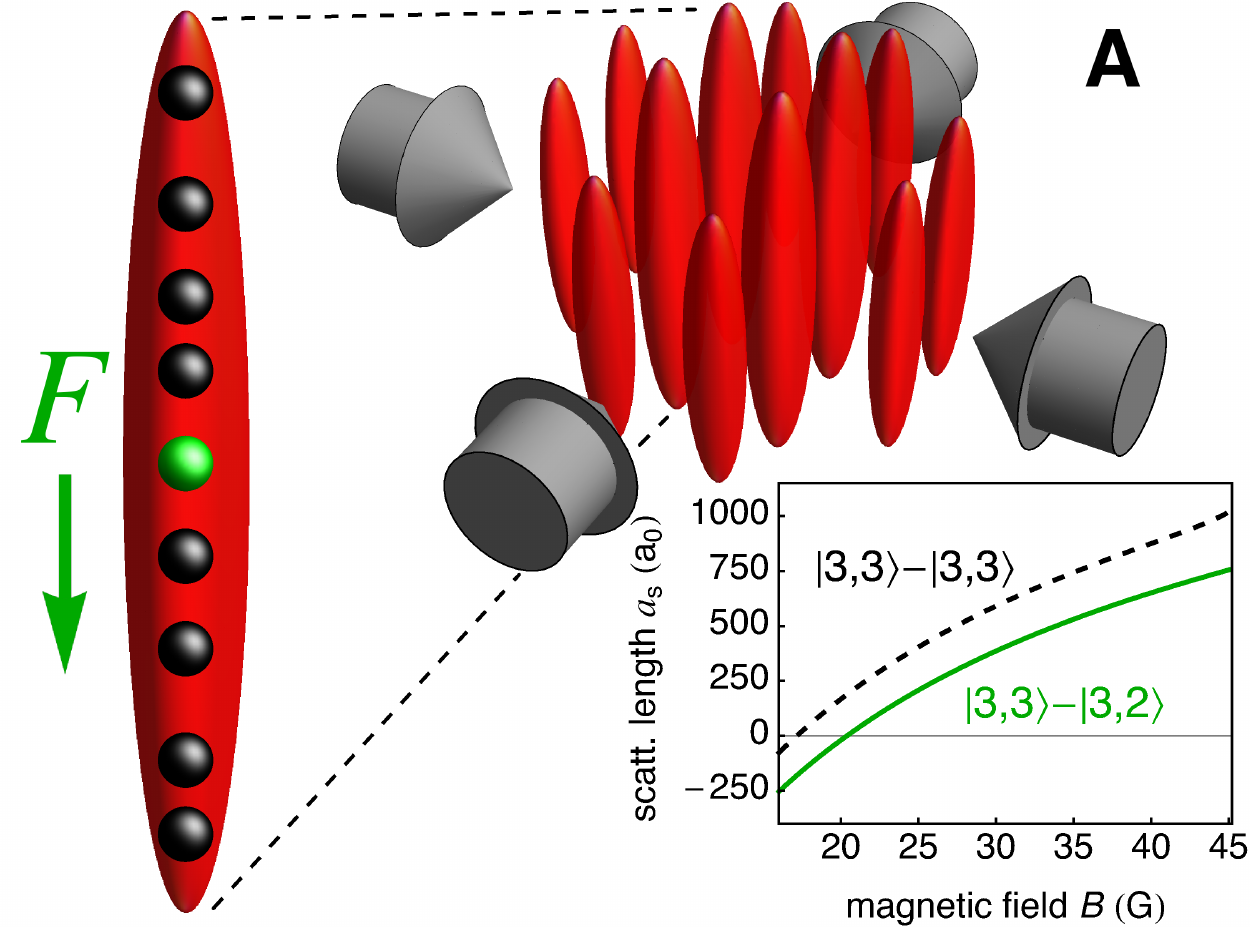}
\hspace{1mm}
\includegraphics[width=0.373\textwidth]{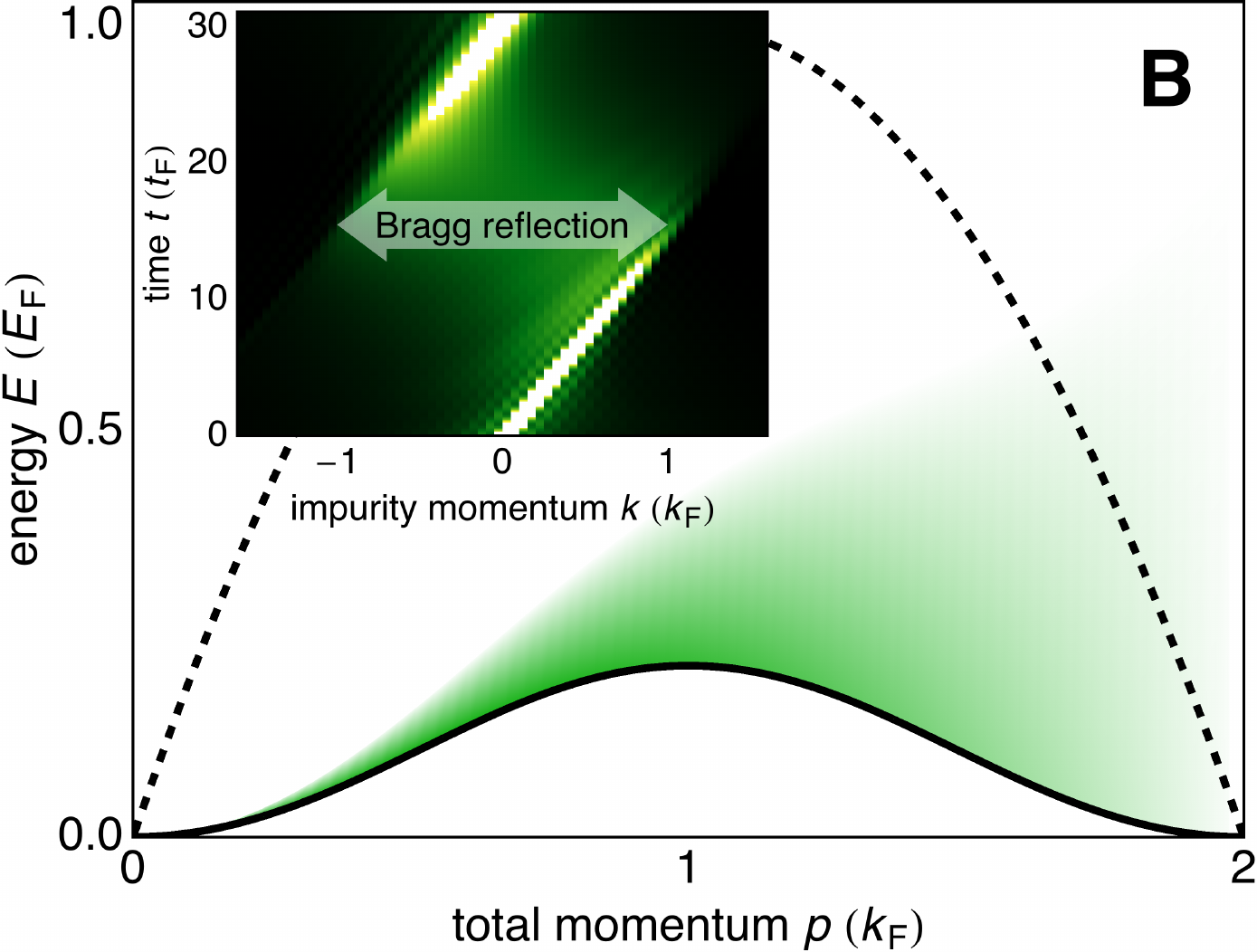}
\caption{\label{FIG1} \textbf{Concept of the experiment.} \textbf{(A)} We realize an ensemble of 1D Bose gases in tubes formed by two pairs of counter-propagating and interfering laser beams. In each tube, a single strongly interacting impurity (green sphere) is immersed in the correlated host gas (black spheres) and is accelerated by gravity (green arrow). Inset: Scattering length $a_{\rm{s}}$ for collisions between the atoms in the host gas (dashed line) and between the impurity and the host atoms (solid line) as a function of the magnetic field $B$. \textbf{(B)} The excitation spectrum of the impurity coupled to the 1D Bose liquid is a $2k_{\rm{F}}$ periodic function of the system's total momentum $p$, bounded from below by a spectral edge (solid line). For comparison, the dashed line indicates the lower bound of excitations in the background gas without impurity. A force acting on the impurity gradually increases $p$ and induces a population of the continuous many-body spectrum above the spectral edge due to non-adiabatic scattering processes (green shading). When the impurity approaches the edge of the correlation-induced Brillouin zone ($k=k_{\rm{F}}$) the background gas can absorb excitations with momentum $2k_{\rm{F}}$ without energy cost, which manifests itself in the impurity's momentum distribution by Bragg reflections. Inset: Numerical simulations of a Bragg reflection for infinitely strong background gas interactions $\gamma=\infty$, strong but finite impurity-host interactions $\gamma_{\rm{i}}=12$, and a weak force $\mathcal{F}=1$.}
\end{figure}

\newpage

\begin{figure*}
\includegraphics[width=0.3\textwidth]{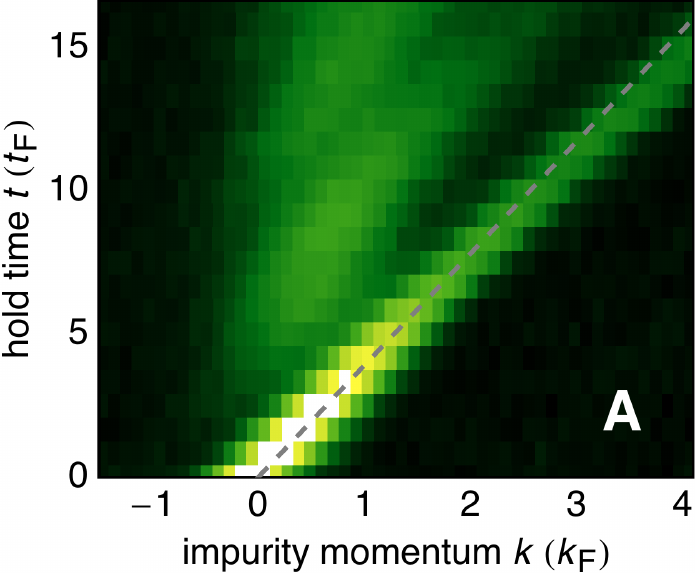}
\hspace{1mm}
\includegraphics[width=0.3\textwidth]{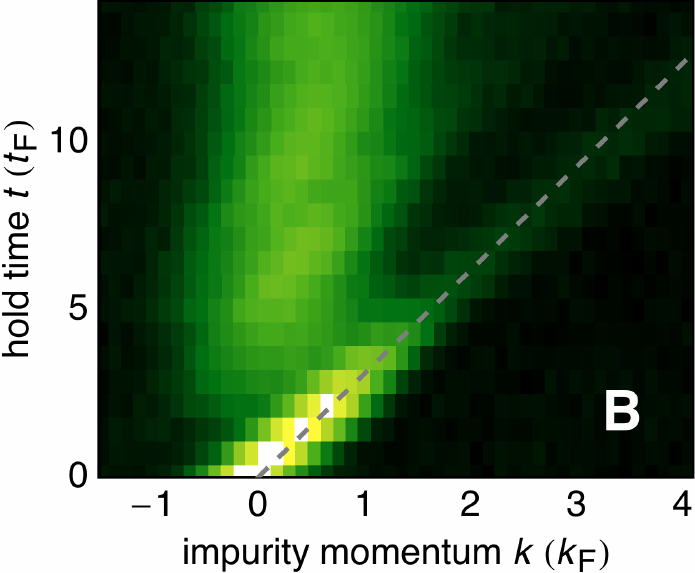}
\hspace{1mm}
\includegraphics[width=0.3\textwidth]{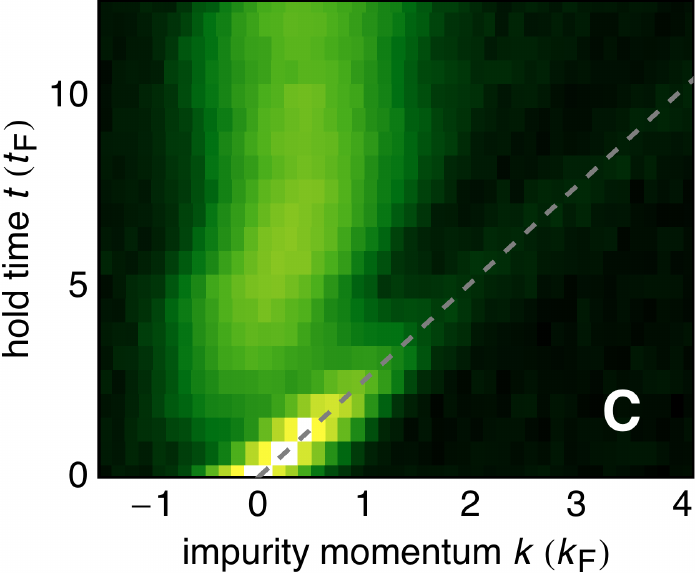}
\hspace{1mm}
\includegraphics[width=0.052\textwidth]{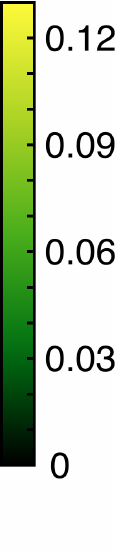}\\
\vspace{1mm}
\includegraphics[width=0.3\textwidth]{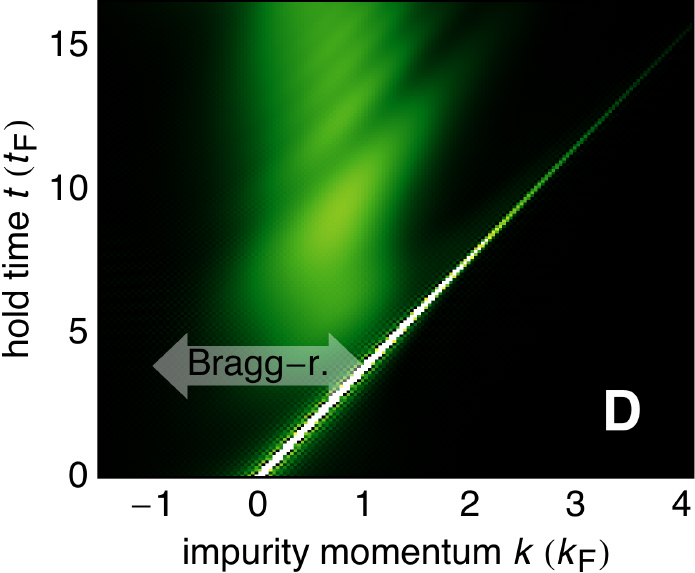}
\hspace{1mm}
\includegraphics[width=0.3\textwidth]{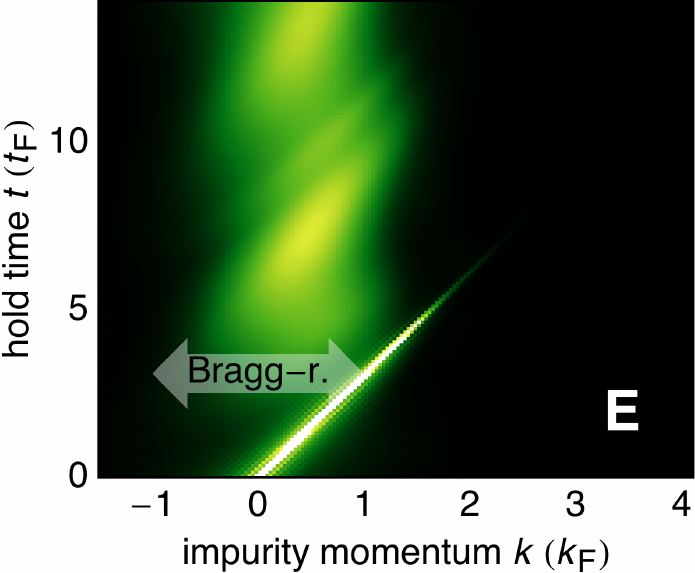}
\hspace{1mm}
\includegraphics[width=0.3\textwidth]{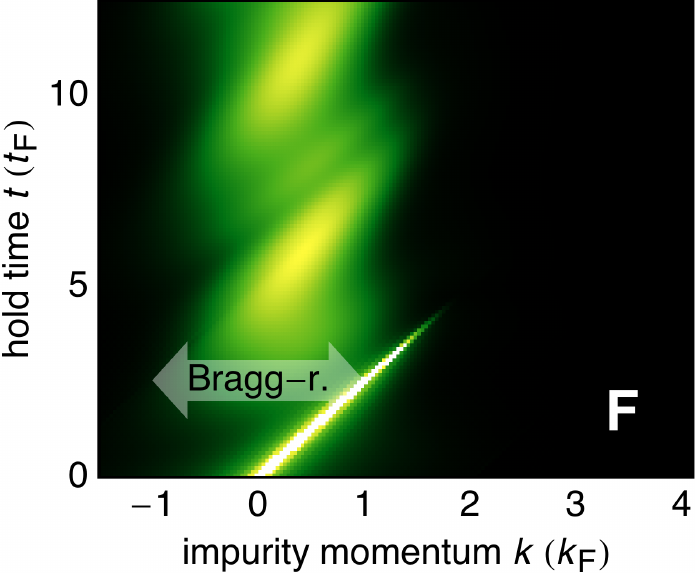}
\hspace{1mm}
\includegraphics[width=0.052\textwidth]{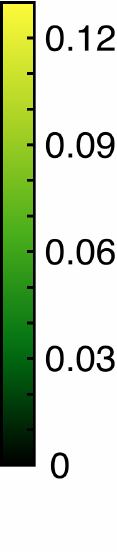}\\
\caption{\label{FIG2} \textbf{Momentum distribution $n(k)$ of the accelerated impurity as a function of time.} Experimental data for $n(k)$ (top row) and numerical simulations (bottom row) are shown for $[\gamma, \gamma_{\rm{i}}] = [7.8(3) , 3.4(1)]$ \textbf{(A},\textbf{D)}, $[15.2(5) , 7.9(3)]$ \textbf{(B},\textbf{E)}, and $[38(1) , 19.4(7)]$ \textbf{(C},\textbf{F)}. In the experiment, the calculated values for $[k_{\rm{F}}$, $t_{\rm{F}}]$ averaged over the sample are $[4.8(2) \, \mu \rm{m}^{-1}$, $0.18(1) \, \rm{ms}]$ \textbf{(A)}, $[4.5(2) \, \mu \rm{m}^{-1}$, $0.21(1) \, \rm{ms}]$ \textbf{(B)}, and $[4.2(1) \, \mu \rm{m}^{-1}$, $0.24(2) \, \rm{ms}]$ \textbf{(C)}. The dashed lines in \textbf{(A)}-\textbf{(C)} indicate the impurity momentum corresponding to a free fall in the residual gravitational field of magnitude $g/3$. Arrows in \textbf{(D)}-\textbf{(F)} depict Bragg reflection of the impurity with momentum transfer $-2 k_{\rm{F}}$. Standard errors given for the experimental parameters reflect a $\pm10$\% uncertainty in the total atom number.}
\end{figure*}
\twocolumngrid

Here, we employ a degenerate gas of Cesium atoms to study the quantum motion of impurities that are immersed in a strongly correlated 1D Bose gas and are accelerated by a constant force. Despite the system being a translational invariant quantum liquid, we observe Bloch oscillations in the impurity dynamics with characteristic Bragg reflections at a wave vector corresponding to the bosonic interparticle distance, which defines an emergent Brillouin zone. We demonstrate that the phenomenon arises for a wide range of system parameters and is robust.

We consider 1D gases of short-range repulsively interacting bosons with mass $m$, prepared in an array of tubes formed by interfering laser beams (\figc{FIG1}{A}). Each 1D system is characterized by the dimensionless parameter $\gamma$~\cite{supmat}. While for $\gamma \ll 1$ the gas is weakly interacting, $\gamma \gg 1$ signifies the strongly correlated Tonks-Girardeau (TG) regime \cite{Girardeau60,Kinoshita04,Haller09}. The impurity immersed in the quantum liquid is of identical mass $m$ and interacts with the host particles with strength $\gamma_{\rm{i}}$. When subject to a force $F$, the dynamics of the impurity is strongly affected by the spectrum of collective excitations of the coupled system. The lower edge of the spectrum is cosine-shaped and periodic, resembling the conventional dispersion of a lattice system (\figc{FIG1}{B}). The existence of the lower edge is a unique feature of 1D and results from kinematic constraints~\cite{Cazalilla11, Zvonarev09}. Furthermore, the excitation spectrum of interacting 1D bosons can be recast in terms of fermions and a Fermi momentum $k_{\rm{F}}$ can be introduced to characterize the periodicity of the spectral edge. As a consequence of Luttinger's theorem \cite{Giamarchi04}, the periodicity is twice the Fermi momentum $2 k_{\rm{F}} = 2 \pi n_{\rm{1D}}$, where $n_{\rm{1D}}$ is the density of host atoms, independent of the interactions. This defines an effective Brillouin zone, which is restricted to momenta between $-k_{\rm{F}}$ and $k_{\rm{F}}$.

The force $F$, characterized by the dimensionless parameter $\mathcal{F} = F m / (\hbar^2 n_{\rm{1D}}^3)$, that acts on the impurity increases the total momentum $p$ of the many-body system linearly in time. Yet, the impurity exhibits Bragg reflections at the edge of the emergent Brillouin zone that change its momentum by $-2 k_{\rm{F}}$ (inset in \figc{FIG1}{B}). The excess momentum is transferred into low-energy particle-hole excitations of the host gas. The resulting oscillatory impurity dynamics can be interpreted as Bloch oscillations. However, this picture is substantially challenged by the gapless continuum of excitations above the spectral edge. These continuum states get excited in the course of the quantum evolution even for weak external forces, which can be understood by a breakdown of adiabaticity in gapless quantum systems. Here, we study experimentally to what extent the oscillatory impurity dynamics can prevail.

Starting from a 3D Cesium Bose-Einstein condensate \cite{Kraemer2004}, we use two pairs of counter-propagating and interfering laser beams to confine the atoms to an array of approximately 3500 vertically oriented and highly elongated 1D systems \cite{supmat}. The combined trapping potentials cause an inhomogeneous atom distribution across the array of tubes with a peak occupancy of about 60 particles. Our confined atoms are initially prepared in their lowest magnetic hyperfine state $|F,m_F\rangle = |3,3\rangle$ and are levitated against gravity by a vertical magnetic field gradient $\nabla B \approx 31.1$ G/cm. We set $\gamma$ by adiabatically raising the scattering length $a_{\rm{s}}$ via an offset magnetic field $B$ using a Feshbach resonance (inset in \figc{FIG1}{A}). The impurity is encoded in a different Zeeman substate $|3,2\rangle$. Applying a short ($50 \mu \rm{s}$) resonant radio-frequency pulse, we create about 3500 impurities in total, i.e., on average a single one per tube. The value of $B$ set prior to the radio-frequency transfer also determines the scattering length $a_{\rm{s}}^{3,2}$ for impurity collisions with the host gas atoms (inset in \figc{FIG1}{A}) and thereby the interaction parameter $\gamma_{\rm{i}}$. Owing to the smaller magnetic moment, the impurity particles are accelerated by one third of gravity $g$. We let the system evolve for a variable hold time $t$ up to $3$ ms before we determine the impurity momentum distribution $n(k)$ in a time-of-flight (TOF) measurement. To this end, the magnetic field is rapidly ramped to $B\approx 21$ G within $50 \mu \rm{s}$. Here, $a_{\rm{s}}^{3,2}$ is sufficiently close to zero that the subsequent motion of the impurities in the tubes is not affected by the host gas atoms. This allows us to reconstruct $n(k)$ from images of the spatially separated spin states after TOF \cite{supmat}.

Results of such measurements are shown in \figc{FIG2}{A-C} for increasing $\gamma$ and $\gamma_{\rm{i}}$, which we determine using the calculated peak density in each 1D system and averaging over the ensemble of tubes \cite{supmat}. Momentum and time are expressed in units of Fermi momentum $k_{\rm{F}}$ and Fermi time $t_{\rm{F}} = \hbar/E_{\rm{F}} = 2m/(\hbar k_{\rm{F}}^2)$, likewise evaluated via an ensemble average. For all data sets we observe initial dynamics consistent with free fall due to the residual gravitational force $F=m g/3$ (dashed line). Yet, as the impurity momentum approaches $k_{\rm{F}}$, a considerable fraction of the impurity is Bragg reflected by $-2 k_{\rm{F}}$ to smaller momenta resulting in clearly bimodal distributions. These atoms again accelerate and a similar second scattering process of weaker contrast is apparent at later times. For stronger interactions the Bragg reflected fraction of the distribution increases (\figc{FIG2}{B}), and deep in the TG regime (\figc{FIG2}{C}), almost the entire ensemble is scattered multiple times to lower momenta.
 
Our experimental findings are supported by numerical simulations based on matrix product states~\cite{Knap14}. We study a single 1D system, described by the many-body Hamiltonian
\begin{equation}
\hat{H} = \hat{H}_{\rm{LL}}(g_{\rm{1D}}, \{ z_n \} ) - \frac{\hbar^2}{2 m} \frac{\partial^2}{\partial {z}^2} + g_{\rm{i}} \sum\limits_{n=1}^{N} \delta (z_n - z) + F z \, .
\label{eq:H}
\end{equation}
Here, $\hat{H}_{\rm{LL}}$ denotes the Lieb-Liniger Hamiltonian accounting for the host gas of $N$ bosons with coordinates $z_n$, interacting via repulsive contact interactions of strength $g_{\rm{1D}}= \hbar^2 n_{\rm{1D}} \gamma /m$ \cite{supmat}. The remaining terms describe the accelerated impurity interacting with all host particles ($g_{\rm{i}}= \hbar^2 n_{\rm{1D}} \gamma_{\rm{i}} /m$). The impurity is initially injected into a gas of $N=60$ particles in its equilibrated lowest momentum state, before we switch on the force $F$. The ensuing time evolution of the impurity momentum distribution (\fig{FIG2}{D-F}) agrees remarkably well with the experiment, including the characteristic Bragg reflections. The fact that the measurements can be understood by a single 1D system demonstrates the robustness and universality of the observed phenomenon. We also identify slight quantitative differences. First, the fraction of $n(k)$ that falls freely through the host gas is systematically larger in the experiment than in the theoretical data. We attribute this to contributions from outer tubes filled with only a few particles or impurities that are initially excited near the lower edge of the ensemble. Thereby these realizations are less affected by the host atoms. Second, the idealized situation with the impurity initiated in its ground state is not fully realized experimentally, as the dynamics starts immediately after its creation. Third, the initial spread of the experimentally measured $n(k)$ is larger compared to the theoretical predictions, which can be mostly attributed to the finite time of flight that sets the measurement resolution~\cite{supmat}.

\begin{figure}
\includegraphics[width=0.47\columnwidth]{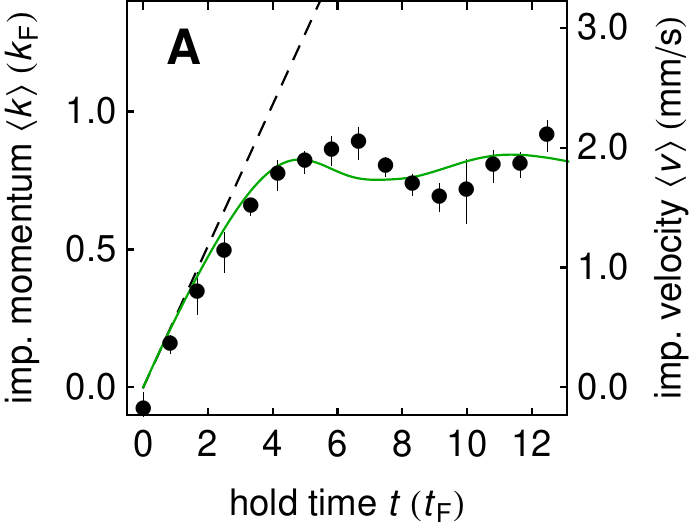}
\hspace{1mm}
\includegraphics[width=0.47\columnwidth]{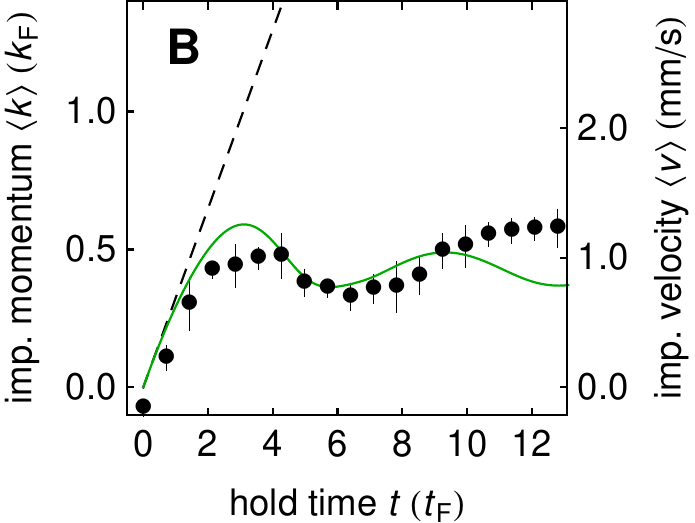}\\
\vspace{1mm}
\includegraphics[width=0.47\columnwidth]{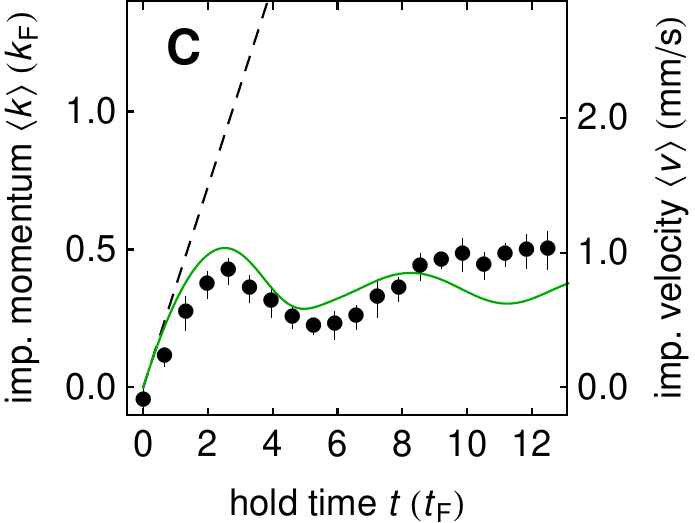}
\hspace{1mm}
\includegraphics[width=0.47\columnwidth]{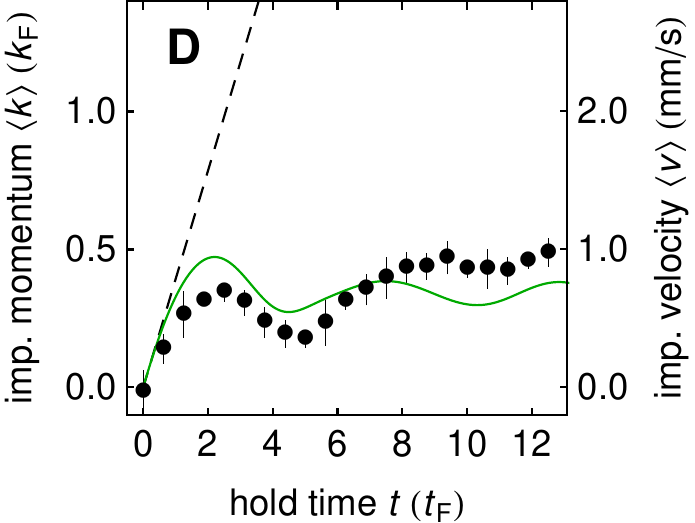}
\caption{\label{FIG3} \textbf{Mean impurity momentum $\langle k \rangle$ as a function of time.} The interaction parameters are $[\gamma, \gamma_{\rm{i}}] = [7.8(3) , 3.4(1)]$ \textbf{(A)}, $[15.2(5) , 7.9(3)]$ \textbf{(B)}, $[24.7(9) , 13.2(5)]$ \textbf{(C)}, and $[38(1) , 19.4(7)]$ \textbf{(D)}. Error bars indicate the standard deviation extracted from typically five realizations per datapoint. Solid lines show the result of numerical simulations and dashed lines the momentum of a free falling impurity.}
\end{figure}

As a consequence of the observed Bragg reflections in $n(k)$, we find that the mean impurity momentum $\langle k \rangle$ exhibits clear damped oscillations before the impurity approaches a finite drift momentum $\langle k \rangle_{\rm{d}}$ (\fig{FIG3}). We compare the time evolution of the measured impurity momentum with our simulations, and find good agreement for different values of $\gamma$ and $\gamma_{\rm{i}}$. In order to obtain the experimental datapoints for the mean impurity momentum we evaluate $n(k)$ in the interval $k \leq |2 k_{\rm{F}}|$ spanning two effective Brillouin zones, which removes the contribution of residual free falling impurities due to the experimental sample inhomogeneity~\cite{supmat}. The chosen momentum range essentially comprises the full simulated momentum distribution.

Our simulations indicate that during the time evolution, excitations are generated in the scattering continuum above the spectral edge even at weak external force. This manifests in energy that is continuously dissipated in the system at a rate that increases with increasing $\mathcal{F}$. As a consequence, $n(k)$ broadens as time evolves (\fig{FIG2}) and the Bloch oscillating impurity experiences damping toward finite drift velocities (\fig{FIG3}). Remarkably, the drift is characterized by a constant rate of dissipated energy, reminiscent of Joule heating \cite{supmat}.

\begin{figure}
\includegraphics[width=0.47\columnwidth]{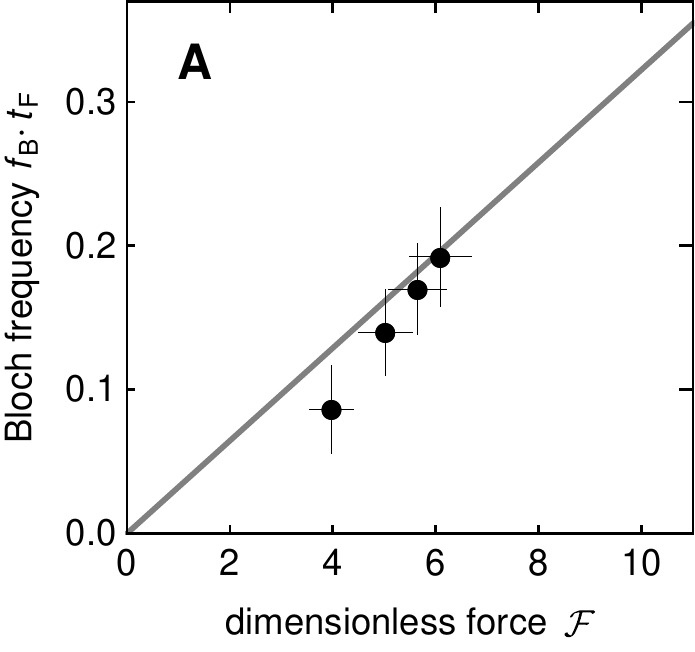}
\hspace{1mm}
\includegraphics[width=0.47\columnwidth]{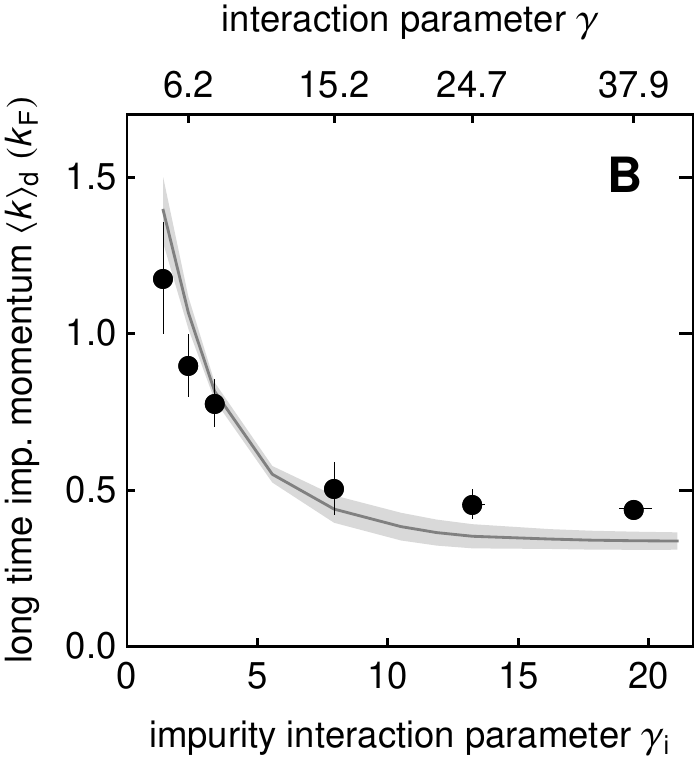}
\caption{\label{FIG4} \textbf{Bloch oscillation frequency and drift momentum.} \textbf{(A)} Measured Bloch oscillation frequency $f_{\rm{B}}$ multiplied by the Fermi time $t_{\rm{F}}$ as a function of the dimensionless force $\mathcal{F}$. The solid line shows the linear dependence $\mathcal{F}/\pi^3$ predicted from a model which considers the $2 k_{\rm{F}}$ periodicity of the spectral edge. \textbf{(B)} Drift momentum $\langle k \rangle_{\rm{d}}$ of the impurity as a function of interactions $\gamma_{\rm{i}}, \gamma$. The solid line shows the prediction extracted from numerical simulations and the shaded region indicates an estimated error mainly resulting from residual oscillations.}
\end{figure}

Additional insight can be obtained from investigating the Bloch oscillation frequency $f_{\rm{B}}$ and drift momentum $\langle k \rangle_{\rm{d}}$. We obtain $f_{\rm{B}}$ by Fourier analyzing the measured mean impurity momentum $\langle k \rangle (t)$ in Fig.~\ref{FIG3}. The results are depicted in \figc{FIG4}{A} and compared with the prediction $f_{\rm{B}} t_{\rm{F}} =  F t_{\rm{F}}/(2 \hbar k_{\rm{F}})  = \mathcal{F} / \pi^3$, which only takes into account the $2 k_{\rm{F}}$ periodicity of the spectral edge. This simple model holds for the experimentally probed forces, as we also verified numerically. The long-time drift momentum $\langle k \rangle_{\rm{d}}$, extracted by taking the average of $\langle k \rangle(t)$ in the range $8 \leq t/t_{\rm{F}} \leq 13$ to remove the residual oscillatory behavior, is shown in \figc{FIG4}{B} as a function of $\gamma_{\rm{i}}$. The significant reduction of $\langle k \rangle_{\rm{d}}$ with increasing coupling strength is a result of stronger Bragg reflections of the impurity with increasing interactions.

Our observations demonstrate a striking dynamical phenomenon, arising from the interplay of strong quantum correlations and far-from-equilibrium conditions. The controlled realization of transient many-body states marked by a high degree of coherence counters the simplest perspective of fast relaxation predicted by hydrodynamics. Moreover, our experiments emphasize the importance of coherence on transport in correlated quantum matter \cite{Laird15} and support that coherence can play an important role in far-from-equilibrium dynamics of many-body systems, as recently discussed in the context of physical, chemical, and biological systems~\cite{Mathy12, Knap14, Romero14,Hiscocka16}. Apart from that, our results reveal how a mobile impurity can probe many-body effects, which here are crystalline correlations resulting from the fermionized nature of the interacting 1D Bose gas. Controlling impurity dynamics holds future prospects for exploring exotic properties of quantum matter, including fractional quantum Hall states and associated topological invariants \cite{Grusdt15} or Anderson's orthogonality catastrophe \cite{Cetina16,Knap12}.

We are indebted to R. Grimm for generous support. We thank O. Gamayun, O. Lychkovskiy, C. Mathy, and M. Schecter for fruitful discussions, P. S. Julienne for providing scattering length data, and E. Haller for contributions in the early stage of the experiment. We gratefully acknowledge funding by the European Research Council (ERC) under Project No. 278417, the Austrian Science Foundation (FWF) under Project No. I1789-N20, and the Technical University of Munich - Institute for Advanced Study, funded by the German Excellence Initiative and the European Union FP7 under grant agreement 291763. E. D. acknowledges support from Harvard-MIT CUA, NSF Grant No. DMR-1308435, AFOSR Quantum Simulation MURI, the Humboldt Foundation, and the Max Planck Institute for Quantum Optics.

\bibliographystyle{Apsrev}

\newpage
\clearpage
\appendix

\setcounter{figure}{0}
\setcounter{equation}{0}

\renewcommand{\thepage}{S\arabic{page}} 
\renewcommand{\thesection}{S\arabic{section}} 
\renewcommand{\thetable}{S\arabic{table}}  
\renewcommand{\thefigure}{S\arabic{figure}} 
\renewcommand{\theequation}{S\arabic{equation}} 

\section{Supplemental Material: Bloch oscillations in the absence of a lattice}

\subsection{Preparation of the array of 1D Bose gases}

Our experiment starts with a Bose-Einstein condensate (BEC) without detectable non-condensed fraction of typically $1.1 \times 10^5$ Cs atoms in the energetically lowest hyperfine ground state $|F,m_F\rangle = |3,3\rangle$ confined in a crossed beam optical dipole trap. Trapping and cooling procedures are described in Refs.~\cite{Weber2002MAT,Kraemer2004MAT}. The sample is levitated against gravity by means of a magnetic field gradient $\nabla B \approx 31.1$ G/cm oriented along the vertical $z$-direction. To prepare the array of 1D Bose gases, we adiabatically load the BEC (within $500$~ms) into an optical lattice generated from two orthogonal and horizontally propagating retro-reflected laser beams at a wavelength $\lambda=1064.5$ nm. After the loading procedure the lattice depth in each direction $x$ and $y$ is $V_{x,y} = 25 E_{\rm{R}}$, where $E_{\rm{R}} = h \times 1.325$ kHz denotes the photon recoil energy for Cs atoms associated with the lattice wavelength and $h$ is Planck's constant. The atoms are then confined to an array of approximately 3500 vertically oriented 1D systems with a transversal trap frequency $\omega_\perp/(2\pi) = 13.2$ kHz. The weak longitudinal confinement caused by the combined trapping potentials is measured to $\omega_z/(2\pi) = 17.9(0.1)$ Hz, giving an aspect ratio of more than 700. In the lattice, we set the interaction strength $\gamma$ by adiabatically raising the scattering length $a_s$ exploiting a broad magnetic Feshbach resonance with a pole at $\sim -12$~G \cite{Mark2011MAT,Haller09MAT}. The ramp time (50 ms) is chosen carefully, i.e. slow enough to avoid any excitations of breathing modes in the gas. 

\subsection{Derivation of experimental system parameters}

In our experiment, the 1D coupling constant $g_{\rm{1D}}$ quantifying interactions in the host gas is set by the scattering length $a_{\rm{s}}$ via \cite{Haller09MAT,Olshanii98MAT}
\begin{equation}
g_{\rm{1D}} = 2 \hbar \omega_\perp a_{\rm{s}} \left( 1- 1.0326 \frac{a_{\rm{s}}}{a_\perp} \right)^{-1} \, .
\end{equation}
Here, $a_\perp = \sqrt{\hbar / (m \omega_\perp)}$ is the radial oscillator length. Analogously, the scattering length for collisions between impurity and host atoms $a_{\rm{s}}^{3,2}$ sets the coupling constant $g_{\rm{i}}$.

The derivation of the dimensionless interaction strengths $\gamma$ and $\gamma_{\rm{i}}$ further requires knowledge of the host gas density. To this end, we first deduce the distribution of the atomic sample across the array of 1D tubes. Assuming sufficiently small interactions during the adiabatic lattice loading, so that all tubes are in the 1D Thomas-Fermi regime, the atom number $N_{i,j}$ for the tube ($i,j$) can be derived iteratively from the global chemical potential (see \cite{Haller11MAT,Meinert15MAT} for details).

For each tube, we then model the 1D density distribution $n(z)$ individually by numerically solving the Lieb-Liniger system and making a local density approximation \cite{Lieb63MAT,Dunjko01MAT}. The derived peak 1D density in each tube $n_{i,j}^{\rm{1D}}$ together with $N_{i,j}$ provides the local and ensemble averaged values for $\gamma$, $\gamma_{\rm{i}}$ and $k_{\rm{F}}$
\begin{subequations}
\begin{equation}
\gamma_{i,j} = \frac{m g_{\rm{1D}}}{\hbar^2 n_{i,j}^{\rm{1D}}} \, , \qquad \gamma = \frac{1}{N} \sum\limits_{i,j} N_{i,j} \gamma_{i,j} \, ,
\end{equation}
\begin{equation}
\gamma^{\rm{i}}_{i,j} = \frac{m g_{\rm{i}}}{\hbar^2 n_{i,j}^{\rm{1D}}} \, , \qquad \gamma_{\rm{i}} = \frac{1}{N} \sum\limits_{i,j} N_{i,j} \gamma^{\rm{i}}_{i,j} \, ,
\end{equation}
and
\begin{equation}
k_{\rm{F}}^{i,j} = \pi n_{i,j}^{\rm{1D}} \, , \qquad k_{\rm{F}} = \frac{1}{N} \sum\limits_{i,j} N_{i,j} k_{\rm{F}}^{i,j} \, .
\end{equation}
\end{subequations}
Here, $N=\sum_{i,j} N_{i,j}$ denotes the total number of atoms. The average Fermi time $t_{\rm{F}}$ and the dimensionless force $\mathcal{F}$ is derived from $k_{\rm{F}}$ via $t_{\rm{F}}=2m/(\hbar k_{\rm{F}}^2)$ and $\mathcal{F} = F m / (\hbar^2 (k_{\rm{F}}/\pi)^3)$, respectively.

\subsection{Radio frequency transfer}

The presence of the magnetic field gradient $\nabla B$ causes a $z$-dependent resonance frequency for the transfer of host gas atoms into the state $|F,m_F \rangle = |3,2\rangle$. This allows for spatially selective radio-frequency addressing and thereby for the creation of an impurity wavepacket of width $\sigma_z$ that is initially localized in space. A small $\sigma_{z}$, however, implies a corresponding uncertainty limited spread in momentum space $\sigma_{k}$. For our experiment we estimate $\sigma_z \sim 6.5 \mu \rm{m}$ from the length of the radio-frequency pulse and the width of its associated Fourier power spectrum. The corresponding spread in momentum space $\sigma_{k} = 1/(2 \sigma_{z}) < 0.05 k_{\rm{F}}$. The observed initial width of $n(k)$ in Fig.~2 of the main text is larger and varies in the range $0.22 k_{\rm{F}} \leq \sigma_{\rm{k}} \leq 0.33 k_{\rm{F}}$. This can be largely attributed to the impurities' initial spatial extent in combination with the finite time-of-flight.

\subsection{Momentum distribution and mean momentum of the impurity}

As briefly discussed in the main text, we determine the momentum distribution of the impurity atoms via an in-tube Stern-Gerlach separation of the two magnetic Zeeman substates $|F,m_F \rangle = |3,3 \rangle$ (host gas atoms) and $|3,2 \rangle$ (impurities). In more detail, after the variable hold time $t$ during which the impurities are accelerated and interact with their host liquids, we quickly set the scattering length $a_{\rm{s}}^{3,2}$ sufficiently close to zero so that any subsequent impurity dynamics is now solely determined by gravity. Note that the host liquid remains levitated against gravity. We wait for a time-of-flight $t_{\rm{TOF}} = t_{\rm{tot}} - t$ before switching off all trapping potentials and taking an in-situ absorption image. Here, $t_{\rm{tot}} = 18 \, {\rm{ms}}$ is kept constant for all measurements. The image delivers the real space density distribution $n_{\rm{TOF}}(z)$ of the impurity atoms, where the spatial coordinate $z$ is measured from the center of the host gas. From this, we reconstruct the momentum distribution $n_{\rm{TOF}}(k)$ for each time $t$, by connecting real space and momentum space via simple kinematics
\begin{equation}
k = \frac{m}{\hbar} \left( z - 1/2 \, g/3 \, (t_{\rm{tot}} - t)^2 \right) / \left( t_{\rm{tot}} - t \right) \, .
\end{equation}
This method holds for a sufficiently long time-of-flight so that small initial displacements in real space $z_0$ of the impurity acquired during the interaction time $t$ are negligible compared to $z$. For our experimental parameters, we estimate an upper bound for the ratio $z_0/z \lesssim 0.03$ from the case of a non-intercating impurity in free fall. Finally, we rescale $k$ and $t$ in units of $k_{\rm{F}}$ and $t_{\rm{F}}$ and normalize the momentum distribution for a direct comparison with theory. As discussed above, the initial width of the measured momentum distribution can be largely attributed to the impurities' initial spread in real space, which essentially limits our experimental resolution in $k$-space.

\begin{figure}
\includegraphics[width=0.47\columnwidth]{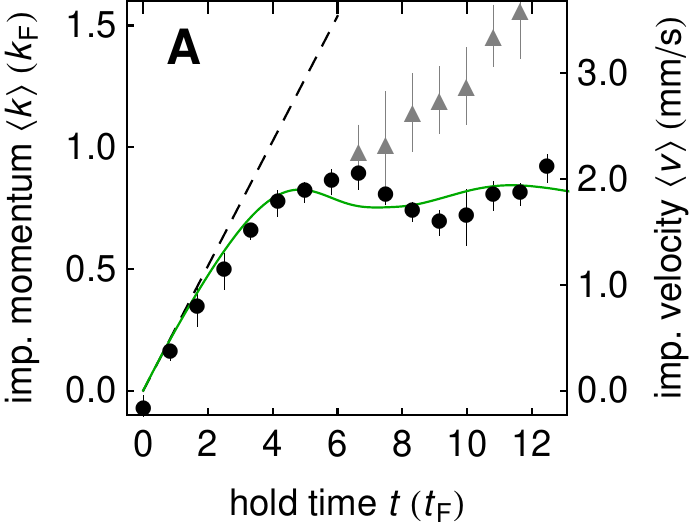}
\hspace{1mm}
\includegraphics[width=0.47\columnwidth]{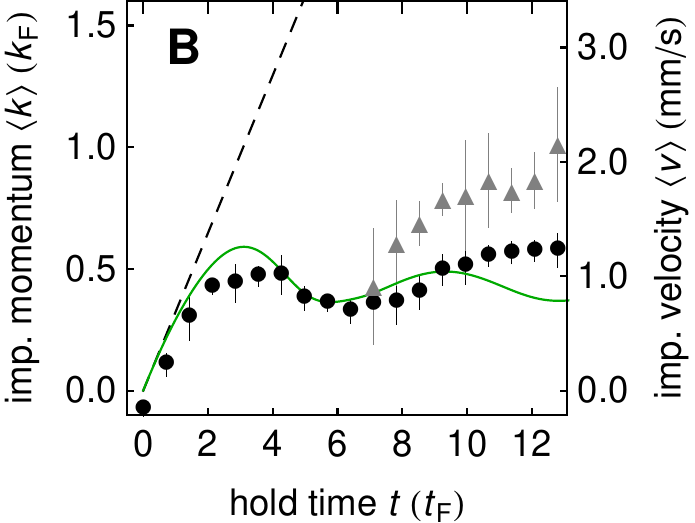}\\
\vspace{1mm}
\includegraphics[width=0.47\columnwidth]{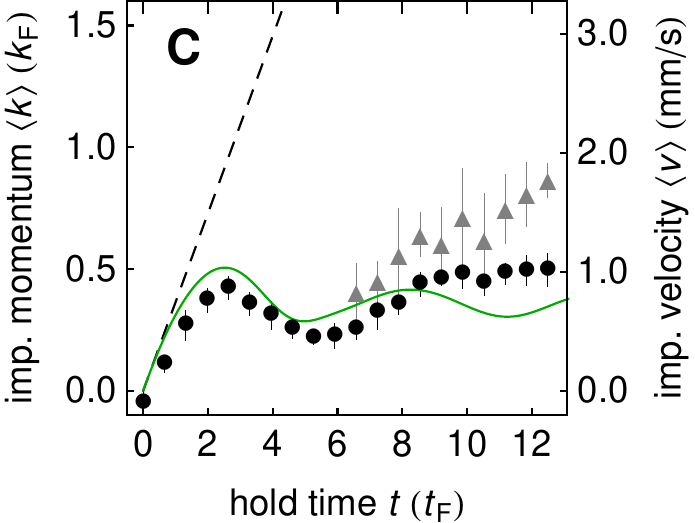}
\hspace{1mm}
\includegraphics[width=0.47\columnwidth]{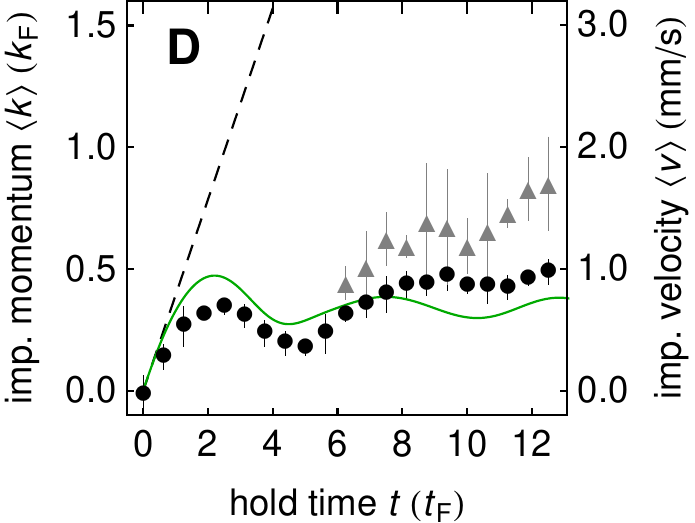}
\caption{\label{SupMatFIG1} \textbf{Extracting the mean impurity momentum.} The mean impurity momentum $\langle k \rangle$ is shown as a function of time for $[\gamma, \gamma_{\rm{i}}] = [7.8(3) , 3.4(1)]$ \textbf{(A)}, $[15.2(5) , 7.9(3)]$ \textbf{(B)}, $[24.7(9) , 13.2(5)]$ \textbf{(C)}, and $[38(1) , 19.4(7)]$ \textbf{(D)}. Circles show the experimental data presented in the main text (\textit{cf.} Fig.~3), obtained by evaluating the measured momentum distribution $n(k)$ of the impurity over two effective Brillouin zones $- 2 \leq k/k_{\rm{F}} \leq 2$. Triangles are obtained from the full measured range of $n(k)$. Solid lines show the result of numerical simulations and dashed lines the free falling impurity.}
\end{figure}

We extract the mean impurity momentum $\langle k \rangle$ from the measured momentum distribution $n(k)$. As we have briefly noted in the main text, we evaluate $n(k)$ in the restricted region $-2 k_{\rm{F}} \leq k \leq 2 k_{\rm{F}}$ comprising two effective Brillouin zones, which essentially contains the numerical data. This procedure allows us to omit residual contribution from free falling impurities, which arises from sample inhomogeneities. In \fig{SupMatFIG1}, we compare the data shown in Fig.~3 of the main article (circles) with results obtained for $\langle k \rangle$ when taking the full measured range of $n(k)$ (triangles). We note that for times $t/t_{\rm{F}} \lesssim 6$ the full-range analysis yields identical results within error bars (not shown), and only at later times, the contribution from free falling impurities has a significant influence.

\subsection{Experimental data for a non-interacting impurity}

\begin{figure}
\includegraphics[width=0.6\columnwidth]{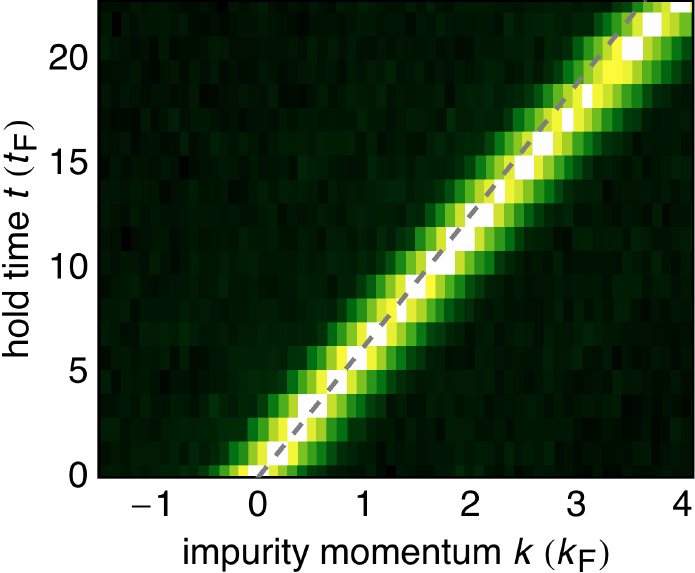}
\hspace{1mm}
\includegraphics[width=0.0501\textwidth]{Impurity1DBO_FIG2G.pdf}
\caption{\label{SupMatFIG2} \textbf{Free fall of a non-interacting impurity.} Experimental data for $n(k)$ as time evolves taken at a magnetic field $B = 20.96(4)$~G, for which the host gas is essentially transparent to the impurity, i.e., $\gamma_{\rm{i}} \sim 0$. The interaction strength characterizing the host gas is $\gamma = 3.0(1)$, the Fermi momentum $k_{\rm{F}} = 5.6(2) \, \mu \rm{m}^{-1}$, and the Fermi time $t_{\rm{F}} = 0.13(1) \, \rm{ms}$. The dashed line indicates the free fall in the residual gravitational field of magnitude $g/3$.}
\end{figure}

As discussed above, the in-tube Stern-Gerlach separation employed for determining $n(k)$ relies on the potential to switch off interactions between the impurity and the host gas during time-of-flight. Here, we demonstrate this capability via a measurement of the dynamics for essentially non-interacting impurities realized at a magnetic field $B \approx 21$~G (\textit{cf.} inset in Fig.~1A in the main text). The obtained $n(k)$ as time evolves, depicted in \fig{SupMatFIG2}, features a clean realization of a free falling impurity in the residual gravitational potential of magnitude $g/3$ without any detectable interactions with the host gas particles.
\begin{figure*}
\includegraphics[width=0.3\textwidth]{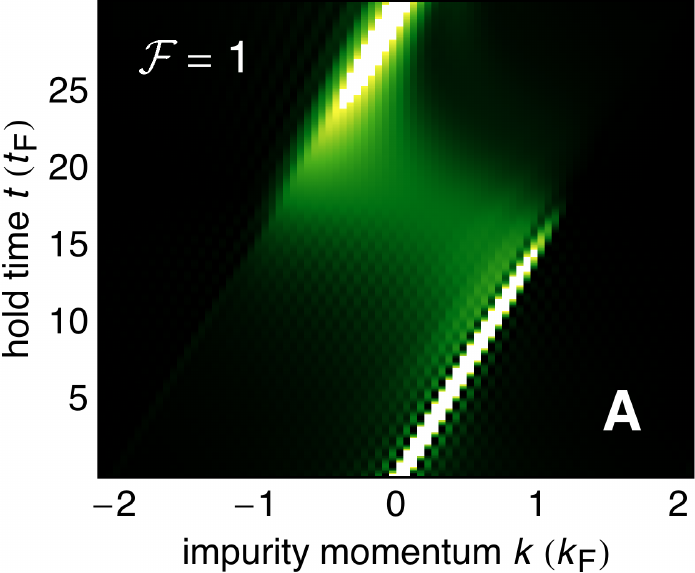}
\hspace{1mm}
\includegraphics[width=0.3\textwidth]{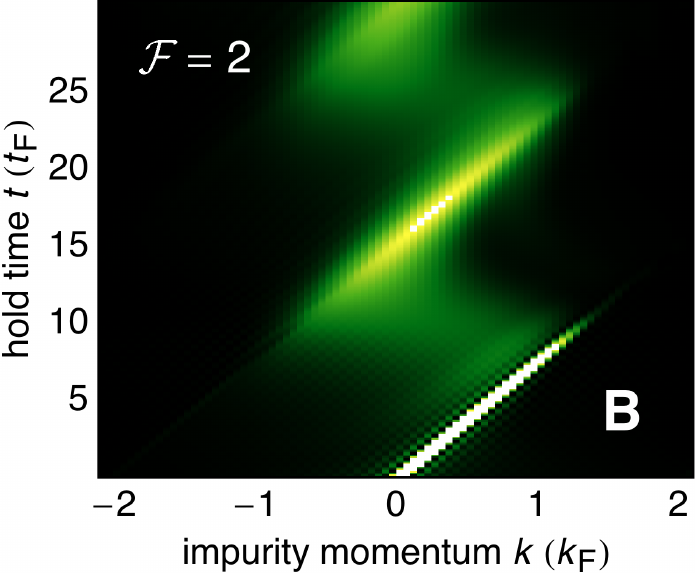}
\hspace{1mm}
\includegraphics[width=0.3\textwidth]{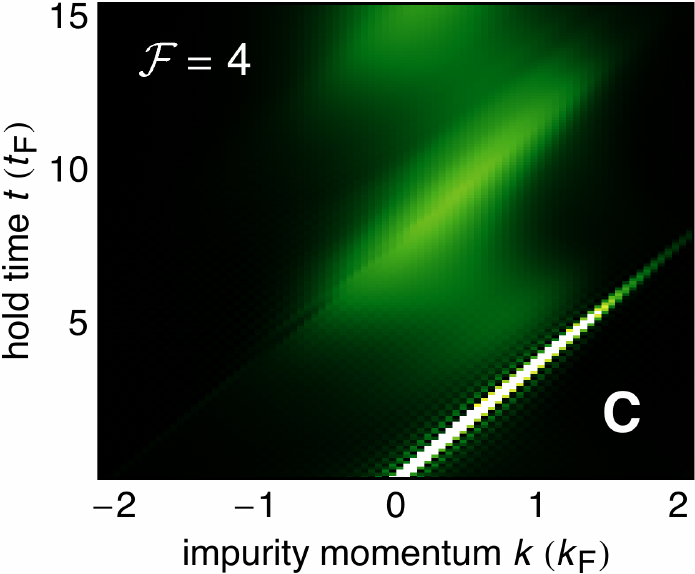}\\
\vspace{1mm}
\includegraphics[width=0.3\textwidth]{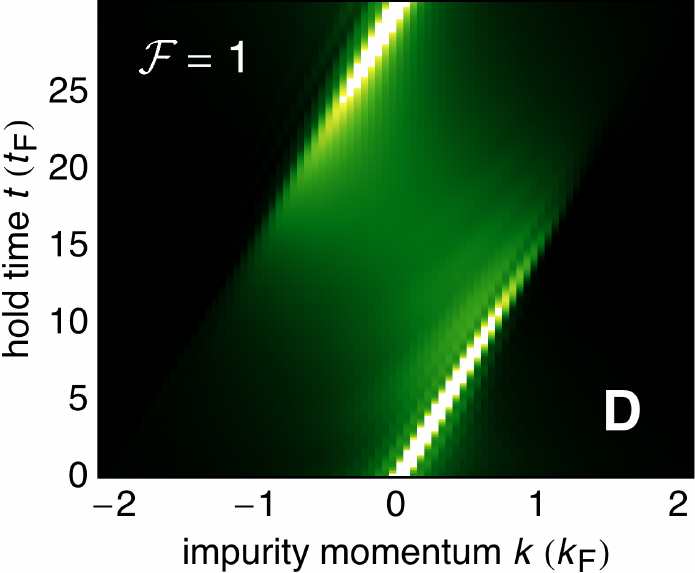}
\hspace{1mm}
\includegraphics[width=0.3\textwidth]{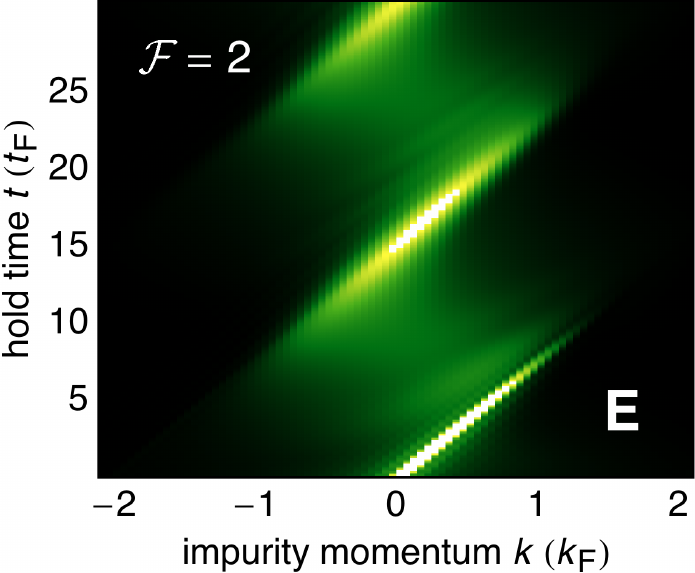}
\hspace{1mm}
\includegraphics[width=0.3\textwidth]{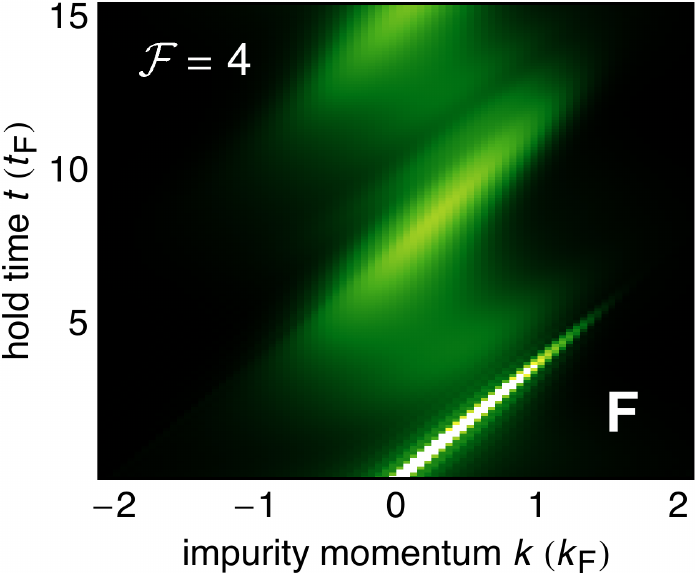}\\
\caption{\label{SupMatFIG3} \textbf{Simulated momentum distribution $n(k)$ of the accelerated impurity as a function of time for increasing force.} Numerical simulations for $n(k)$ are shown for $\mathcal{F}=1$ \textbf{(A},\textbf{D)}, $\mathcal{F}=2$ \textbf{(B},\textbf{E)}, and $\mathcal{F}=4$ \textbf{(C},\textbf{F)}. The data shown in \textbf{(A)}-\textbf{(C)} are for $\gamma_{\rm{i}} = 6$, while \textbf{(D)}-\textbf{(F)} depict results for stronger impurity host coupling $\gamma_{\rm{i}} = 24$. For all panels, the host gas is in the Tonks-Girardeau limit $\gamma = \infty$. Note the different scale on the time axis in panels (C) and (F).}
\end{figure*}

\subsection{Theoretical model}

The full theoretical model [Eq. (1) in the main text] for a single 1D system is
\begin{equation}
\hat{H} = \hat{H}_{\rm{LL}}(g_{\rm{1D}}, \{ z_n \} ) - \frac{\hbar^2}{2 m} \frac{\partial^2}{\partial {z}^2} + g_{\rm{i}} \sum\limits_{n=1}^{N} \delta (z_n - z) + F z \, .
 \label{eq:HS}
\end{equation}
with the Lieb-Liniger Hamiltonian~\cite{Lieb63MAT}
\begin{equation}
 \hat{H}_{\rm{LL}}(g_{\rm{1D}}, \{ z_n \} ) = - \sum_{n=1}^N \frac{\hbar^2}{2 m}\frac{\partial^2}{\partial {z_n}^2} + g_{\rm{1D}} \sum_{1 \leq n < m \leq N} \delta(z_n-z_m). 
 \label{eq:LL}
\end{equation}
The Lieb-Liniger Hamiltonian describes the host gas of $N$ bosons with coordinates $z_n$, interacting via repulsive contact interactions of strength $g_{\rm{1D}}= \hbar^2 n_{\rm{1D}} \gamma /m$. The other terms in the full system Hamiltonian $\hat H$ describe the kinetic energy of the impurity, the interactions of the impurity with the host particles ($g_{\rm{i}}= \hbar^2 n_{\rm{1D}} \gamma_{\rm{i}} /m$), and the linearly increasing potential set by the constant force $F$.

We initiate our system in the state $|\mathrm{in}\rangle$ and evolve it in time
\begin{equation}
|\mathrm{in}(t)\rangle = e^{-i \hat H t / \hbar} |\mathrm{in}\rangle,
\end{equation}
where $\hat H$ is given in \eq{eq:HS}. The impurity momentum distribution $n(k)$ is defined as
\begin{equation}
n(k) = \langle\mathrm{in}(t)| c^\dagger_{k} c^\nag_{k} |\mathrm{in}(t)\rangle,
\end{equation}
where $c^\dagger_{k}$ ($c_{k}^\nag$) creates (annihilates) the impurity with the momentum $k$. The average impurity momentum $\langle k \rangle$ is defined through $n(k)$ as
\begin{equation}
\langle k \rangle = \int dk \, k \, n(k).
\end{equation}
Since the state $|\mathrm{in}(t)\rangle$ is not an eigenstate of the Hamiltonian \eqw{eq:HS}, both $n(k)$ and $\langle k \rangle$ depend on time.

\subsection{Numerical simulations}

We employ matrix product states to investigate the quantum evolution of our system numerically~\cite{Vidal04MAT,Knap14MAT}. To this end, we discretize space and distribute $N=60$ particles over $600$ sites. First, we compute the ground state of our Hamiltonian, Eq.~(1) in the main text, with switched-off force $F=0$. This ensures that the impurity is equilibrated with the background gas. In the experiment the dynamics is initialized by directly transferring on average one host atom in the impurity state, which leads to a finite initial momentum spread. As we demonstrate in the main text, the qualitative features of the impurity dynamics are not sensitive to these details. Having computed the initial state, we switch on the gravitational force and simulate the ensuing time evolution of the system. In order to obtain high-accuracy results we push the numerical simulations to their limits and consider matrix product states with bond dimension $800$. For all the presented results, we carefully assure that they are representative for the continuum limit and converged over the whole time range. 

\subsection{Role of the force}

In our experiment, the dimensionless force acting on the impurity varies in the range $4 \lesssim \mathcal{F} \lesssim 7$. Here, we numerically investigate the role of the finite force on the impurity motion and further demonstrate the robustness of the Bloch oscillation dynamics over a wide range of system parameters.

\begin{figure}
\includegraphics[width=1\columnwidth]{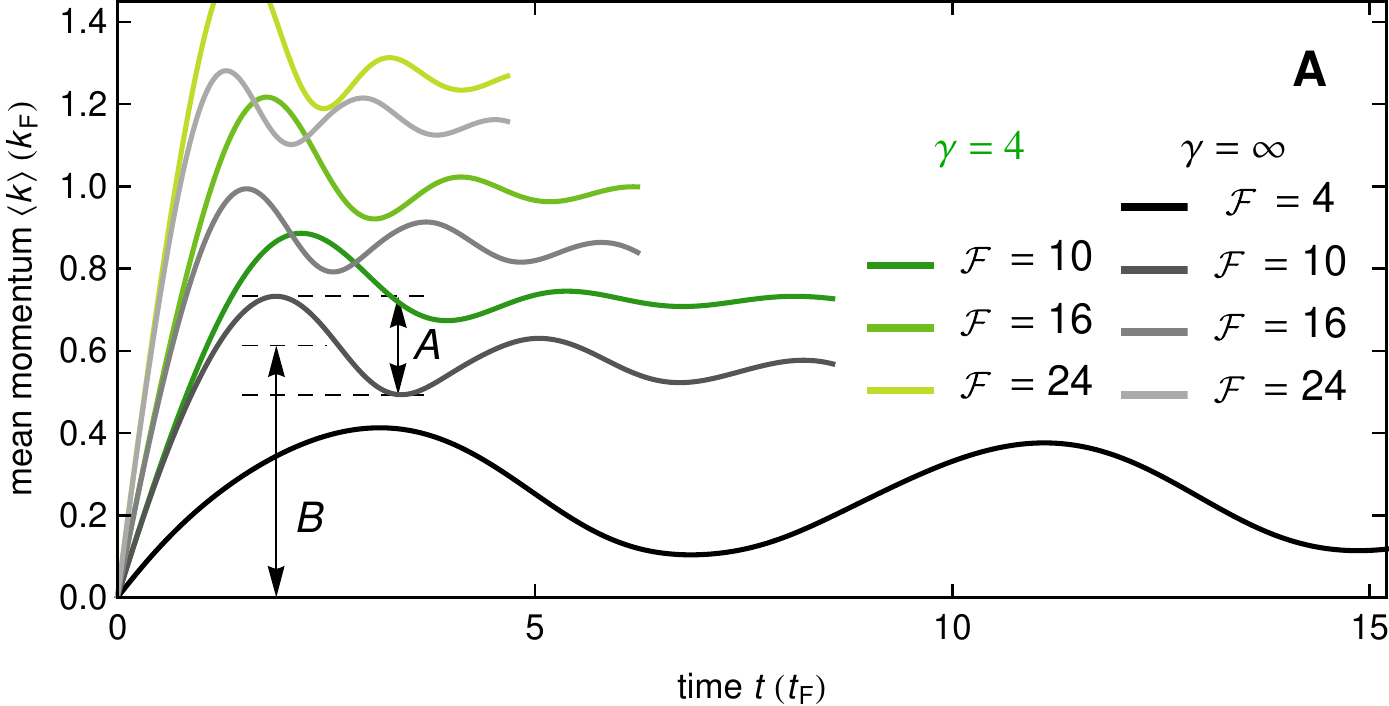}\\
\vspace{2mm}
\includegraphics[width=0.48\columnwidth]{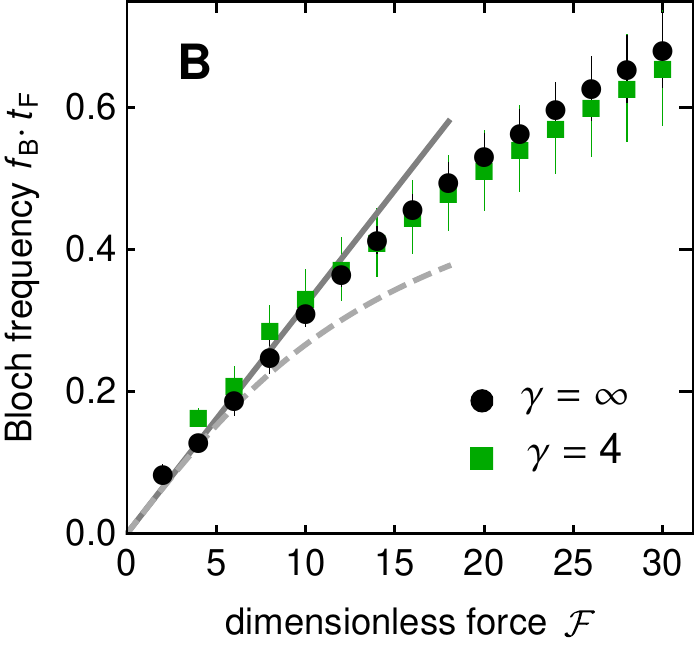}
\hspace{2mm}
\includegraphics[width=0.48\columnwidth]{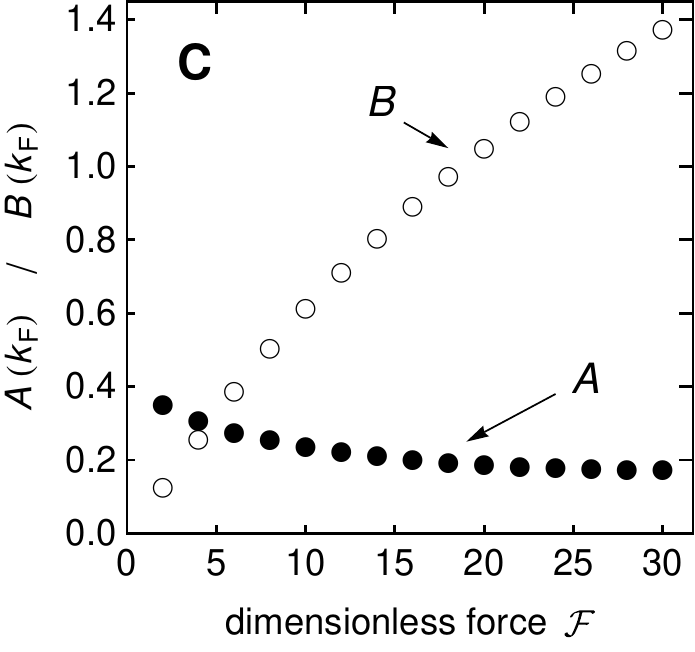}
\caption{\label{SupMatFIG4} \textbf{Numerical simulations for varying force $\mathcal{F}$.} \textbf{(A)} Mean impurity momentum $\langle k \rangle$ as a function of time for $[\gamma, \gamma_{\rm{i}}] = [\infty ,12]$ and $[\gamma, \gamma_{\rm{i}}] = [4 ,12]$ at different values of $\mathcal{F}$. The dashed horizontal lines indicate the minimal momentum $\langle k \rangle_{\rm{min}}$ and maximal momentum $\langle k \rangle_{\rm{max}}$ of the first oscillation extracted to determine $A = \langle k \rangle_{\rm{max}} - \langle k \rangle_{\rm{min}}$ and $B = (\langle k \rangle_{\rm{max}} + \langle k \rangle_{\rm{min}})/2$. \textbf{(B)} Bloch frequency $f_{\rm{B}} \cdot t_{\rm{F}}$ obtained from the numerical data as a function of $\mathcal{F}$ for $[\gamma, \gamma_{\rm{i}}] = [\infty ,12]$ (circles) and $[\gamma, \gamma_{\rm{i}}] = [4 ,12]$ (squares). The solid line shows the linear dependence expected from a simple model taking into account the $2k_\text{F}$ periodicity of the spectral edge. The dashed line indicates the prediction of the model suggested in Ref.~\cite{Schecter12MAT} in the limit $\gamma_{\rm{i}} \gg 1$ and $\gamma=\infty$ neglecting full self-consistency. \textbf{(C)} Numerically extracted values for $A$ (full circles) and $B$ (open circles) as a function of $\mathcal{F}$ for $[\gamma, \gamma_{\rm{i}}] = [\infty ,12]$.}
\end{figure}

We first compute the momentum distribution $n(k)$ for comparatively small forces $\mathcal{F} = 1, \, 2,$ and $4$, for two different values of $\gamma_{\rm{i}}$ while the host gas is in the Tonks-Girardeau (TG) limit, $\gamma= \infty$ (see \fig{SupMatFIG3}). For all parameters, the impurity undergoes oscillatory motion marked by Bragg reflections. Moreover, we find that $n(k)$ broadens as time evolves. The broadening becomes more pronounced as $\mathcal{F}$ increases. Furthermore, a weaker coupling between the impurity and the host particles $\gamma_{\rm{i}}$ shifts the weight in $n(k)$ to larger values. The broadening of $n(k)$ eventually causes relaxation of the oscillatory dynamics toward finite drift velocities. This is most evident, when plotting the mean momentum $\langle k \rangle$ as a function of time for increasing values of $\mathcal{F}$ (\figc{SupMatFIG4}{A}). Our numerical results indicate the persistence of oscillations up to large forces, exceeding the experimental parameters. Also away from the TG limit, the oscillatory motion remains as we demonstrate with additional numerical data shown for a reduced $\gamma=4$.

A shorter oscillation period and a larger drift momentum with increasing $\mathcal{F}$ is also evident from the data. A simple model for the Bloch oscillations, taking into account only the periodicity of the spectral edge, predicts $f_{\rm{B}} t_{\rm{F}} =  F t_{\rm{F}}/(2 \hbar k_{\rm{F}})  = \mathcal{F} / \pi^3$. Our numerical simulations support this model for small forces but indicate deviations to smaller frequencies for larger values of $\mathcal{F}$ (see \figc{SupMatFIG4}{B}). A similar behavior is expected from a model of mobile Josephson junctions  \cite{Schecter12MAT}. 

Finally, let us briefly comment on the relative amplitude of the Bloch oscillations, which we quantify by the peak momenta of the first oscillation cycle, $\langle k \rangle_{\rm{max}}$ and $\langle k \rangle_{\rm{min}}$ (see \figc{SupMatFIG4}{A}). From these values, we determine the amplitude $A$ and the mean value $B$ around which $\langle k \rangle$ oscillates, and show both in \figc{SupMatFIG4}{C} a function of $\mathcal{F}$. Importantly, our experiment is performed in a regime ($\mathcal{F} \lesssim 10$) for which $A$ and $B$ are comparable, thus leading to a sufficiently large relative amplitude of the Bloch oscillations that is well detectable.

\subsection{Dissipated energy}

In the due course of the time evolution, the continuous many-body spectrum of the gapless quantum liquid will be populated, which gives rise to damping toward finite drift velocities $\langle v \rangle_{\rm{d}}$. Here, we investigate the energy dissipated in the system, which provides further insight to what extent the adiabatic following of the lower spectral edge breaks down. Specifically, we evaluate the quantum expectation value of the system Hamiltonian, Eq.~(1) in the main text, for $\mathcal{F}=0$ with the time evolved wave function.

\begin{figure}
\includegraphics[width=1\columnwidth]{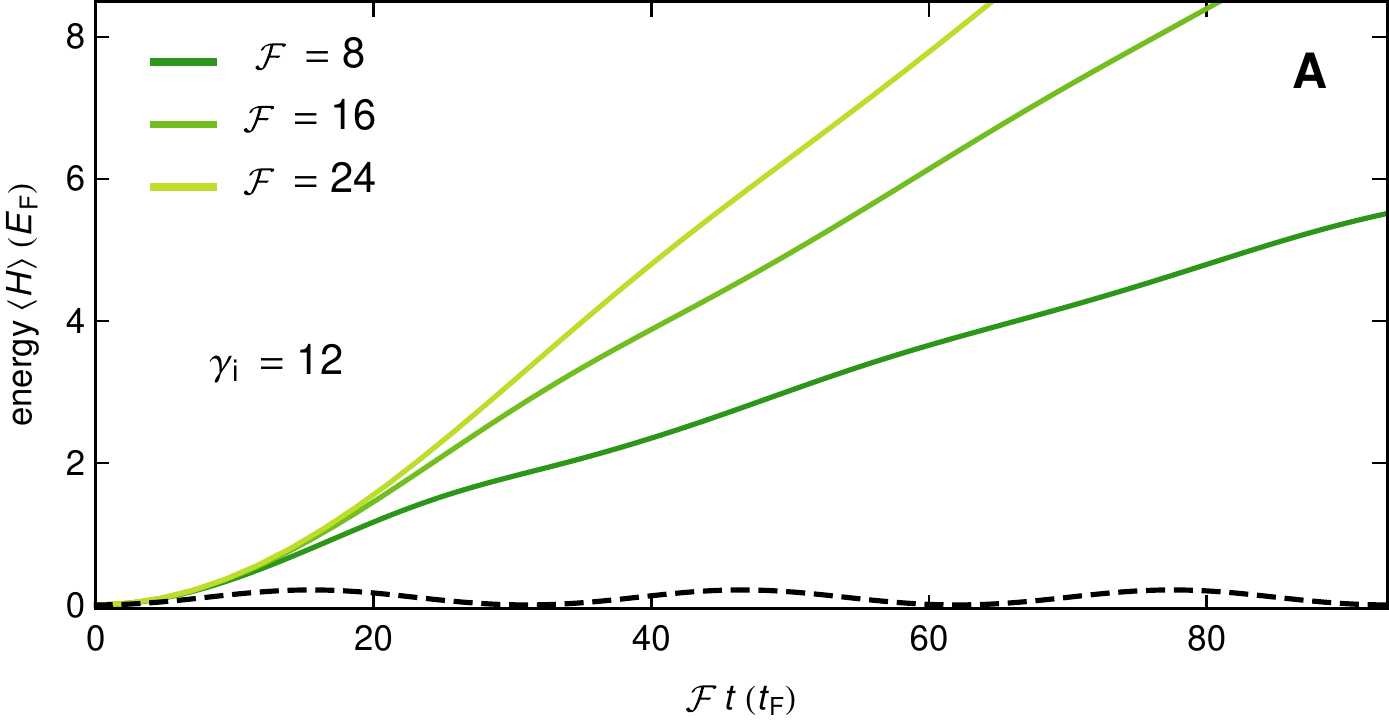}\\
\vspace{2mm}
\includegraphics[width=0.32\columnwidth]{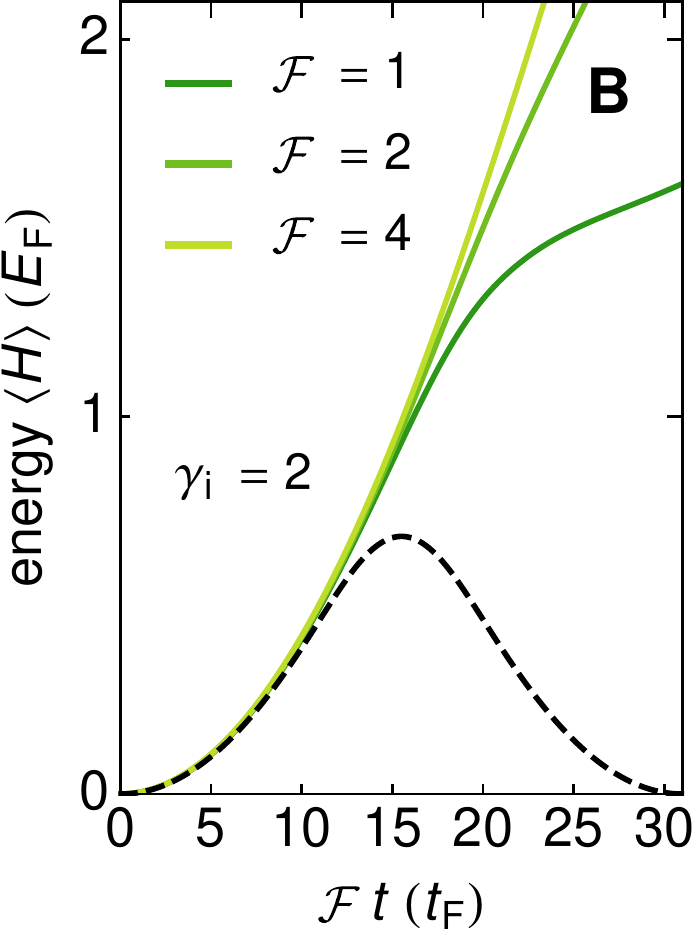}
\hspace{0.2mm}
\includegraphics[width=0.32\columnwidth]{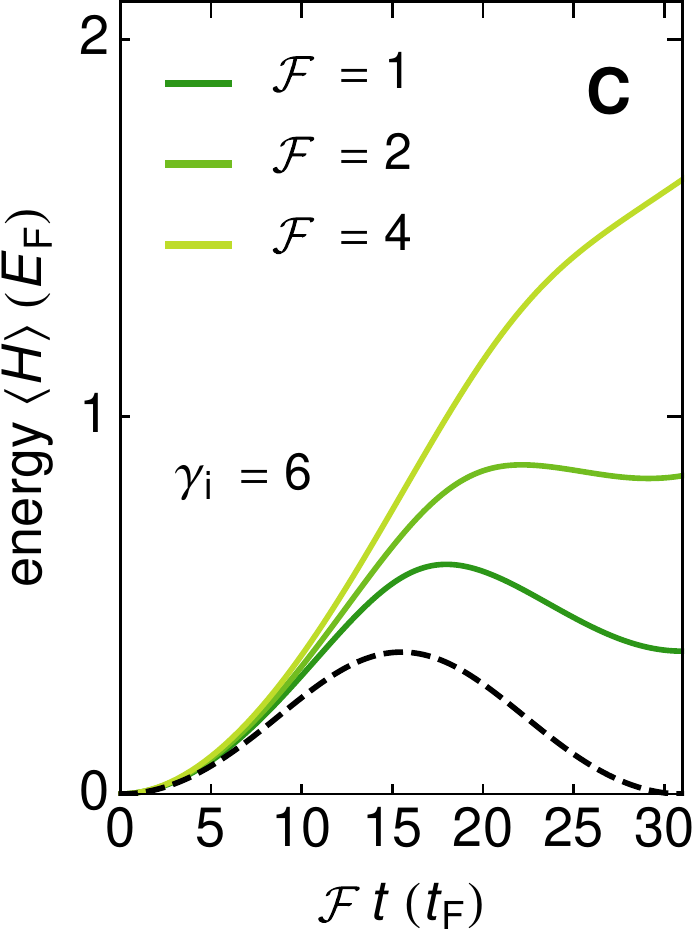}
\hspace{0.2mm}
\includegraphics[width=0.32\columnwidth]{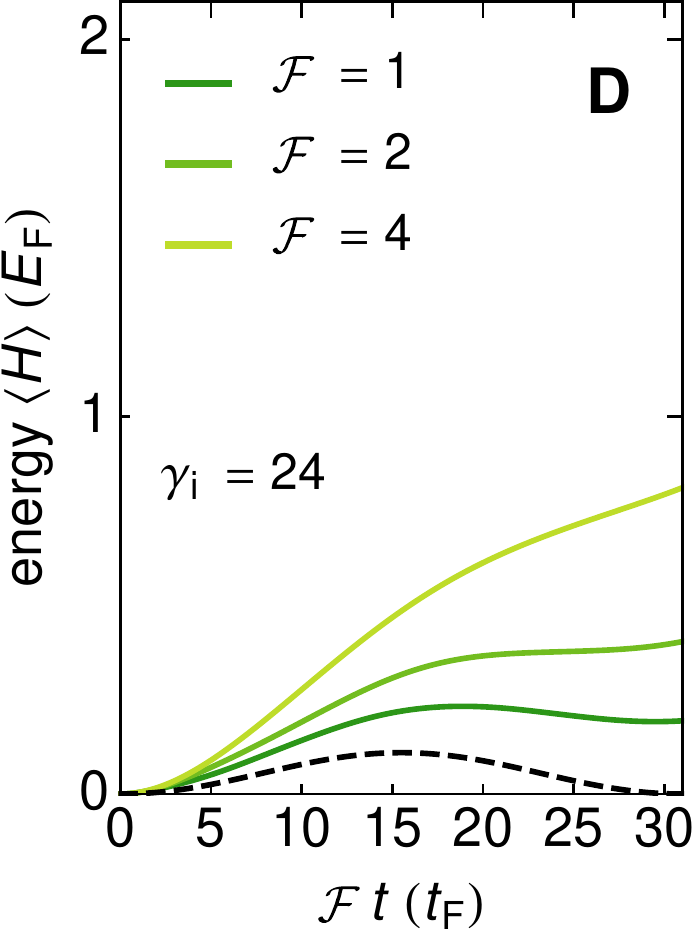}
\caption{\label{SupMatFIG5} \textbf{Numerical simulations of the dissipated energy.} The dissipated energy $\langle H(\mathcal{F}=0) \rangle$ as a function of time multiplied by the dimensionless force $\mathcal{F}$ for strong \textbf{(A)} and weak forces \textbf{(B)} - \textbf{(D)} as indicated in the panels. The impurity-host interaction strength is $\gamma_{\rm{i}} = 12$ in \textbf{(A)}, $\gamma_{\rm{i}} = 2$ in \textbf{(B)}, $\gamma_{\rm{i}} = 6$ in \textbf{(C)}, and $\gamma_{\rm{i}} = 24$ in \textbf{(D)}. For all data $\gamma=\infty$. For comparison, the dashed lines indicate the absorbed energy under the assumption that the system follows adiabatically the lower edge of its excitation spectrum.}
\end{figure}

Numerical results for comparatively strong forces are shown in \figc{SupMatFIG5}{A} for an impurity-host coupling $\gamma_{\rm{i}} = 12$. As time progresses, the dissipated energy increases linearly, reminiscent of Joule heating with a constant power which we find to be consistent with the classical expectation $d\langle H(\mathcal{F}=0) \rangle/dt \sim F \langle v \rangle_{\rm{d}}$.

In \figc{SupMatFIG5}{B-D}, we show further numerical results for weaker forces and three different values of $\gamma_{\rm{i}}$ over the course of the first Bloch-oscillation cycle $0 \leq \mathcal{F} t \leq \pi^3 t_{\rm{F}}$. The observed heating reflects the population of the continuous many-body spectrum above the spectral edge. For sufficiently small $\mathcal{F}$ and sufficiently strong impurity-host coupling, we find a non-monotonic time evolution of the dissipated energy, which approaches the edge of the excitation spectrum. Indeed, the transient decrease of the system's energy as time evolves indicates that a large part of the system's wave function populates states near the spectral edge. These states are responsible for the Bragg reflections of the impurity with the host gas absorbing the excess momentum change without energy cost.

\bibliographystyle{Apsrev}

\newpage
\clearpage

\end{document}